\documentclass[journal]{IEEEtran}
\pdfoutput=1
\usepackage[pdftex]{graphicx}
\usepackage{amsmath}
\usepackage{cite} 

\usepackage[utf8]{inputenc}

\begin{document}

\title{Additive phase-noise in frequency conversion in LLRF systems}

\author{Igor~Rutkowski,
Krzysztof~Czuba
\thanks{Manuscript received June 24, 2018.}
\thanks{I.~Rutkowski and K.~Czuba are with Warsaw University of Technology (WUT), Warsaw, Poland.}
}

\maketitle
\thispagestyle{empty}

\begin{abstract}

This contribution focuses on phase-noise added during frequency conversion in low-level radio frequency (LLRF) control systems. 
The stability of beams’ parameters in linear accelerators depends on the stability of amplitude and phase of the accelerating field. 
A LLRF control system regulates the electromagnetic field inside accelerating modules based on the input RF signals. 
Typically active mixers down-convert those signals, which are later sampled and digitized by ADCs.
This field detection scheme necessitates synthesis of a heterodyne/local oscillator (LO) signal which is often generated using a passive mixer and a frequency divider. 
Additive close–to-carrier phase noise can be observed in the aforementioned circuits.
The phase noise of a passive mixer’s output signal is typically calculated using a small-signal model based on modulation theory. 
Experimental results indicate that the power level of the input signals has a non-linear effect on phase noise beyond the noise floor.
The frequency dividers' output phase noise was measured as a function of input power, input frequency, and division ratio.
The influence of the LO signal power level on the active mixers' output signal phase noise was measured and two hypotheses were made. 
Further measurements of the AM-PM and PM-AM conversion were made to verify one of the hypotheses.
The fidelity of the LO signal is partially determined by the phase noise of the IF signal. 

\end{abstract}

\section{Introduction}

\IEEEPARstart{T}{he} stability of beams’ parameters in linear accelerators depends on the steadiness of amplitude and phase of the accelerating field.
Low-level radio frequency (LLRF) control systems regulate the field using feedforward and feedback loops based on the input RF signals.
Modern digital LLRF control systems typically convert those signals to an intermediate frequency (IF) using active mixers and employ ADCs to sample and digitize IF signals.
The digital logic calculates the output signal, which is later converted back to analog domain and up-converted.
Such field detection scheme necessitates synthesis of a heterodyne/local oscillator (LO) signal for analog frequency conversion as well as a clock signal for ADCs and control logic.
The clock frequency is usually synthesized by a digital frequency divider while the LO frequency is commonly generated using a direct analog frequency synthesis scheme.
Additive phase noise is introduced in all mentioned circuits and directly affects the field stability. 

This paper begins by discussing frequency synthesis schemas commonly used in LLRF control systems.
In Section \ref{s:pn} the current state of research on additive phase noise in investigated devices is summarized and gaps in the current state of knowledge are identified.
New findings are presented in Section \ref{s:meas_modeling}.

\subsection{Frequency synthesizers}
\label{ss:synthesizers}

Synthesizers may be classified into three types:
\begin{itemize}
\item Direct Analog, 
\item Direct Digital,
\item Indirect Digital. 
\end{itemize}
The first approach uses frequency dividers, mixers, and filters. 
In this scheme the output phase noise closely follows that of the reference signal, with additional noise induced by a frequency divider, a mixer, and commonly an amplifier.
Far from carrier the noise can be improved by using a narrow band-pass filter. 
Such filters increase the cost of the device and are sensitive to mechanical vibrations. 

A Direct Digital Synthesizer (DDS) uses a numerically controlled oscillator feeding a digital-to-analog converter, both being synchronized by the reference clock.  
The use of a binary frequency tuning word can result in a systemic frequency error for some intermediate to reference frequency ratios.

An Indirect Digital Synthesizer uses a PLL with an integer or fractional frequency divider. 
For offset frequencies within the loop bandwidth the phase noise is determined by the reference signal’s characteristics distorted by the loop's noise.
Outside the loop bandwidth the noise is dominated  by the VCO used and optional output frequency dividers. 

The Direct Analog scheme is preferred for LLRF systems because no systemic frequency error is introduced and the far from carrier uncorrelated phase noise is reduced.

In the next section, the phase noise in investigated devices is discussed.

\section{Phase noise}
\label{s:pn}

Oscillators generate a periodic signal, which sets the timing of a system.
Phase noise is a representation of random fluctuations in the phase of a signal.
Additive phase noise is the extra jitter added by a component to a signal as it passes through the device.
Phase noise of frequency dividers, active and passive mixers will be investigated.

\subsection{Frequency dividers}
\label{ss:f_div_pn_theory}

Several attempts have been made to qualitatively describe  frequency dividers' phase noise.
Different authors \cite{Kroupa-1982,Phillips-1987,Egan-1990,Levantino-2004,Apostolidou-2008} characterize it as random phase variations, random jitter, or noise power spectral densities.

The published data concerns either ICs which are out-dated and not useful in LLRF control systems  or custom-made ICs which also are not available on the market.
Additionally, manufacturers present data on dividers' phase noise only for selected input frequencies. 
Therefore, an opportunity to advance the state of knowledge exists by characterizing the output phase noise of modern digital frequency dividers in wide ranges of input frequencies and division ratios.


\subsection{Passive mixers}
\label{ss:passive_mixers_theory}

The phase noise of a passive mixer’s output signal is typically calculated using a small-signal model based on the modulation theory \cite{Zuk-2016}.

Barnes \cite{Barnes} measured noise of 18 passive mixers operating as phase detectors (homodyne) for linear and saturated modes of operation.
10 Hz offset frequency was used to characterize the flicker noise and 100 kHz to sample the white noise level.
Many devices performed better when operating in saturation.
In frequency synthesis typically both input ports of the mixer are saturated \cite{Rubiola}, supressing the transmittance of the AM noise.
The maximum difference between flicker noise and thermal noise levels of measured devices was around 40 dB, indicating significant differences between mixers' noise levels can be expected.
No explanation was given for the observed variations between mixers, all of whom were composed of Schottky diodes.

Simulations done by Faber et al. for such diode in the room temperature indicate that the total noise at 1.5 GHz falls with the bias current in the range $10^{-9}$A to $10^{-4}$A and rises above it (see Fig. 3.10, 3.11 and 3.15 in \cite{Faber}).

The analysis above was done for DC-biased operation, in pulsed operation (such as LO signal generation) additional effects must be considered.
The flicker noise is assumed to be a near-dc phenomenon \cite{Rubiola}, which is up-converted to other spectral components.
The thermal noise is constant under LO excitation, but the amplitude of shot noise depends on the diode's current and therefore is pulsed in pumped operation.
Each shot noise spectral component is partially correlated with every other component.
Due to up- and downconversion at each mixing frequency the final shot noise output signal consist of frequency converted components from other frequencies.

According to authors' best knowledge there is no research published on the influence of the input signals' power levels on the IF signal's phase noise in passive mixer operating as heterodyne.
Creation of quantitative description of such relationship is one of this research's goals.  

\subsection{Active mixers}
\label{s:f_div_pn}
Multi-channel field detection modules designed for LLRF systems typically use active mixers to minimize the required LO drive and improve the isolation between channels \cite{Gan-2016}, \cite{Hoffmann-PhD}.
Such mixers use active transistors as non-linear elements.
Often they are combined with amplifiers increasing the strength of one or both of the input signals.
The available literature discusses the noise properties of the Gilbert cell dependent on the MOS transistors' properties.
Early publications \cite{Hull-1993}, \cite{Rudell-1997} focused on thermal and shot noise, assuming the flicker noise to be unimportant \cite{Rofougaran-1996}.
More detailed analysis done by Terrovitis \cite{Terrovitis-1999} concluded that the flicker noise originating in the current-switching MOS transistor pair will appear around DC and it will be transferred only to even-order harmonics of the LO frequency.
It was determined that increasing LO amplitude up to a certain value reduces the noise contribution of the LO port and the switching pair. 
Darabi \cite{Darabi-2000} described the indirect mechanism responsible for flicker noise: changes in the output current of the switching pair due to the capacitance of the tail current generator.  

As mentioned before active mixers are typically equipped with buffer amplifiers that operate in saturation and condition the input signal to be square-wave with amplitude independent of the input signal's level (within certain range).
The datasheet for LT5527, an active mixer IC selected for FLASH's and XFEL's field detection module, states ``The optimum LO drive is –3dBm for LO frequencies above 1.2GHz'' \cite{LT5527_DS}.
This study aims to evaluate the influence of the LO input power on the mixer's output phase noise.

\section{Measurements}
\label{s:meas_modeling}
In previous section the gaps in the state of knowledge were identified.
Measurements were performed to obtain data concerning selected topic of interest.
\subsection{Frequency dividers}
\subsubsection{Component selection}

Many frequency divider ICs are available on the market, but only a very limited subset of them can be used for frequency synthesis in LLRF systems. 
The criteria for selection include phase noise performance and support of division ratios in the tens range.
Some application require the ability to synchronize to an external reset signal.

For comparison four ICs were selected: AD9508, AD9515, HMC7043, LTC6954.
The AD95XX family supports multiple signaling standards, including CMOS, HSTL, LVPECL, and LVDS. 
HMC7043 supports LVDS, LVPECL, as well as CML while LTC6954 supports LVPECL, LVDS, and CMOS.
No information is provided concerning the structure of frequency division circuits in any of the selected devices.
The following high-performance signaling standards were selected for comparison:
\begin{itemize}
\item CML (HMC7043),
\item HSTL (AD908),
\item LVPECL (LTC6954 and AD9515).
\end{itemize}
\subsubsection{Effects of input power level change}

\begin{figure}[htb!]
	\includegraphics[width=\linewidth]{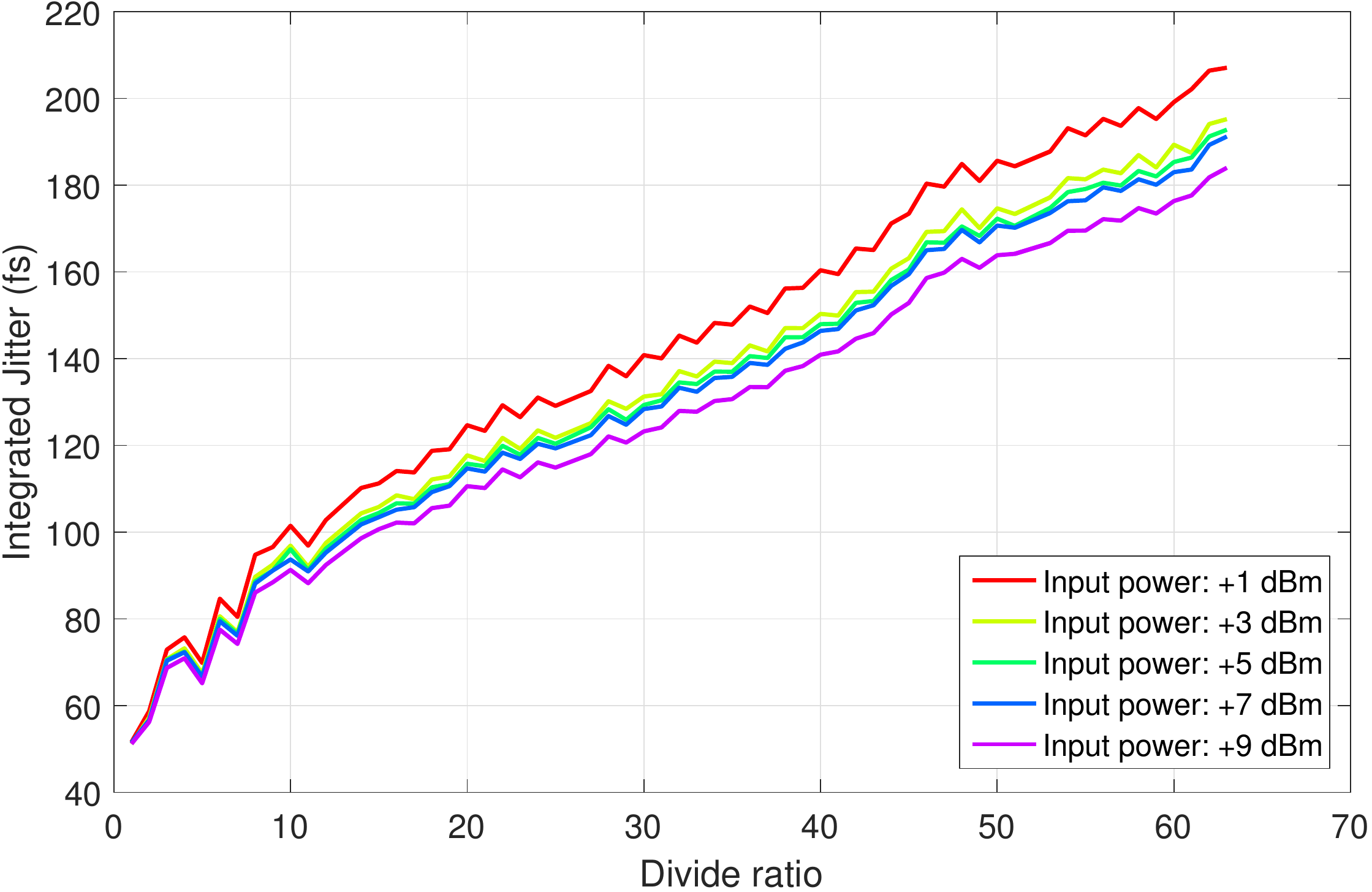}
	\caption{AD9508 (HSTL boost signaling standard) output signal's phase noise integrated in the 10 kHz - 5 MHz band as a function of the division ratio for various input power levels. $f_{in}$ = 1 GHz.}
	\label{fig:ad9508_input_pow}
\end{figure}

\begin{figure}[htb!]
	\includegraphics[width=\linewidth]{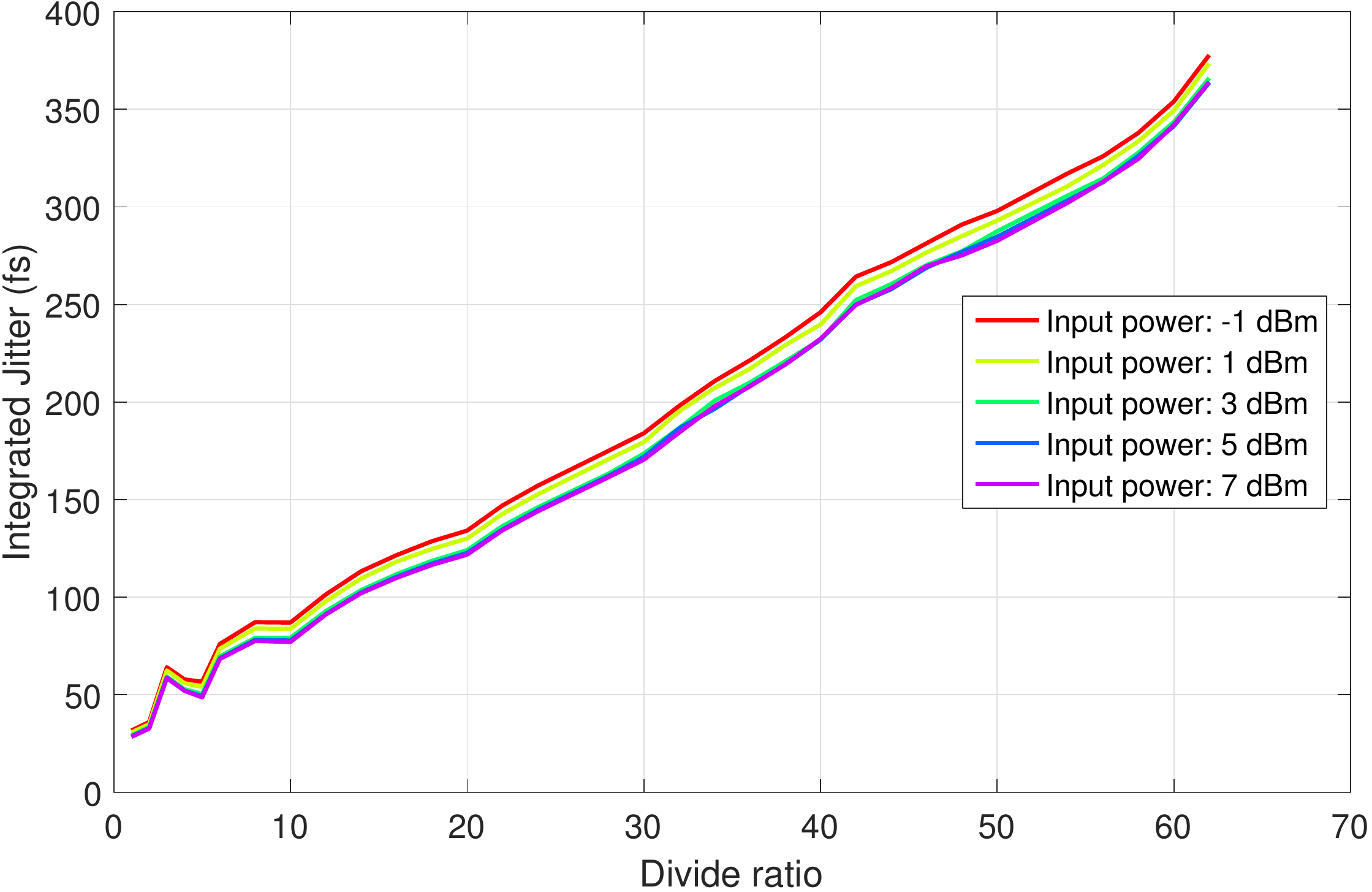}
	\caption{HMC7043 (CML standard) output signal's phase noise integrated in the 10 kHz - 5 MHz band as a function of the division ratio for various input power levels. $f_{in}$ = 1 GHz.}
	\label{fig:hmc7043_lf_input_pow}
\end{figure}

\begin{figure}[htb!]
	\includegraphics[width=\linewidth]{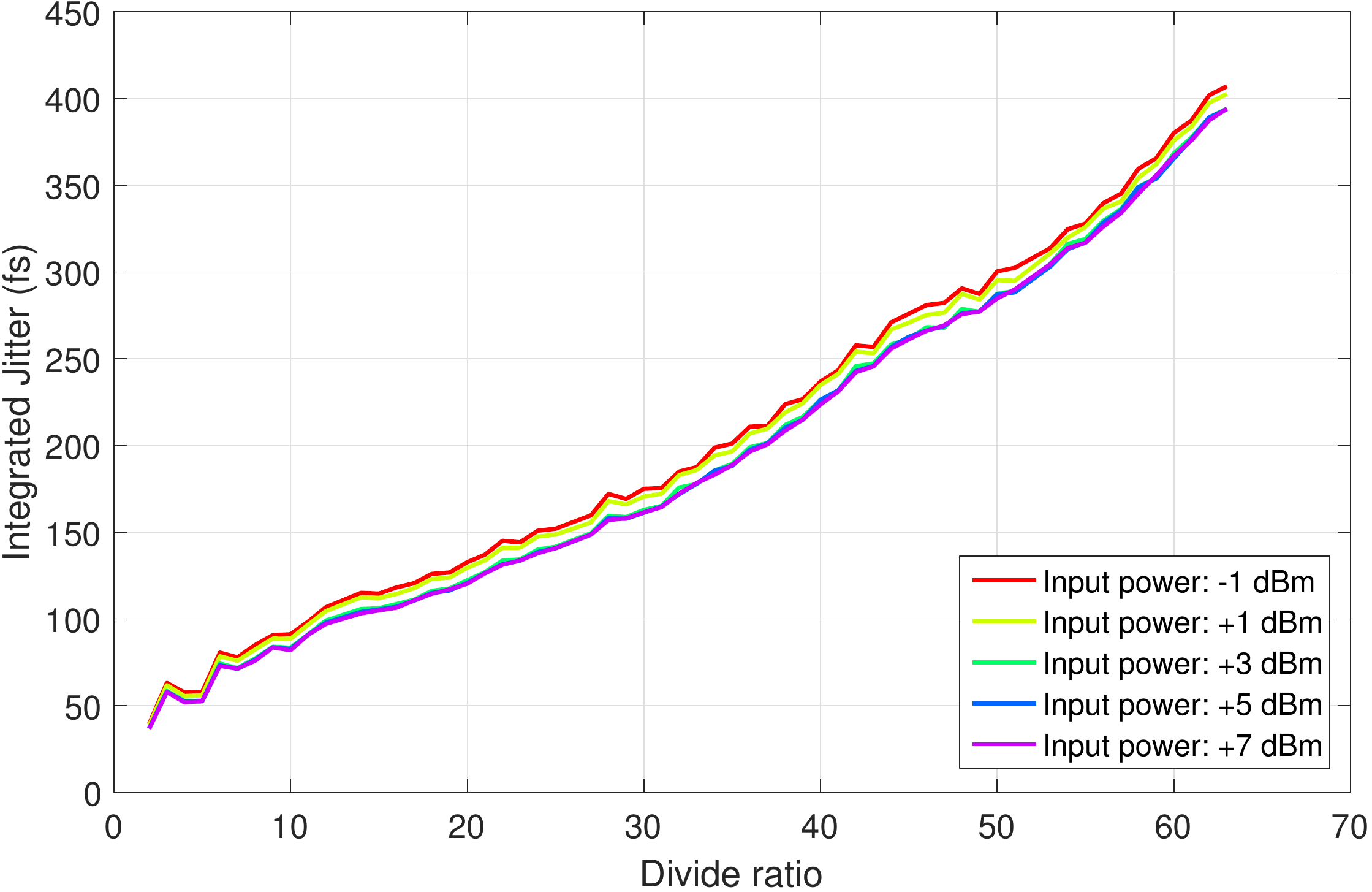}
	\caption{LTC6954 (LVPECL standard) output signal's phase noise integrated in the 10 kHz - 5 MHz band as a function of the division ratio for various input power levels. $f_{in}$ = 1 GHz.}
	\label{fig:ltc6954_input_pow}
\end{figure}

\begin{figure}[htb!]
	\includegraphics[width=\linewidth]{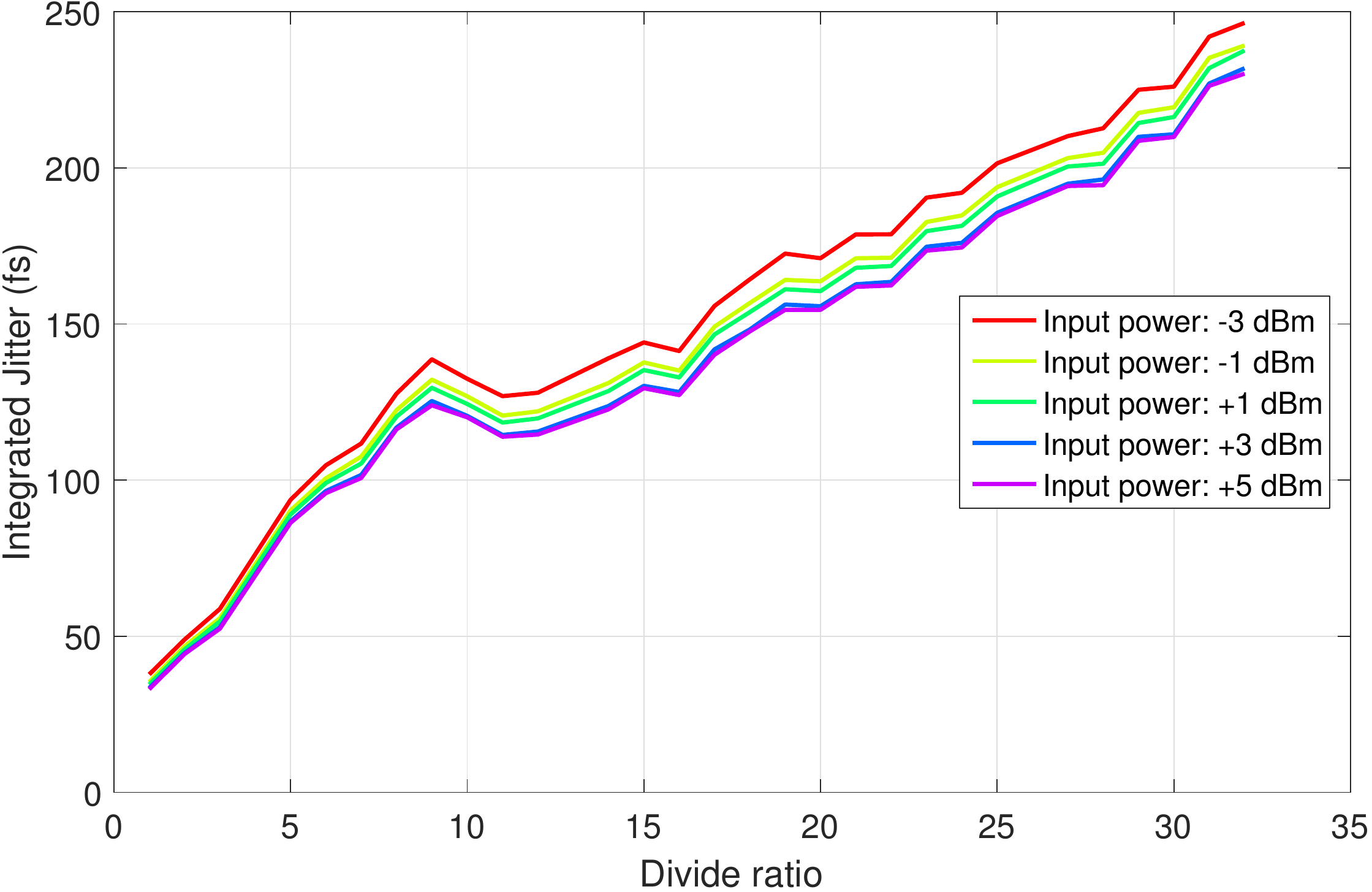}
	\caption{AD9515 (LVPECL 400 mV standard) output signal's phase noise integrated in the 100 Hz - 5 MHz band as a function of the division ratio for various input power levels. $f_{in}$ = 1 GHz.}
	\label{fig:ad9515_input_pow}
\end{figure}

The research began with an investigation of the influence of the input power level on the output phase noise for a given input frequency (1 GHz) and variable divide ratio. 
Agilent E8257D generated the input signal and Agilent E5052B (a Signal Source Analyzer) measured the phase noise.
AD9508, HMC7043 and LTC6954 showed no measurable differences of phase noise in the 10 Hz - 10 kHz band as when the input power level varied (not shown), however in the 10 kHz - 5 MHz band the trend was noticeable (Fig. \ref{fig:ad9508_input_pow}, \ref{fig:hmc7043_lf_input_pow}, and \ref{fig:ltc6954_input_pow}).
In case of AD9515 the influence of the input power level is noticeable above 100 Hz offset frequency (Fig. \ref{fig:ad9515_input_pow}).
For all dividers the jitter falls with the input power level and the spread rises with the division ratio, but input sensitivity differs between ICs.
\begin{figure}[htb!]
	\includegraphics[width=\linewidth]{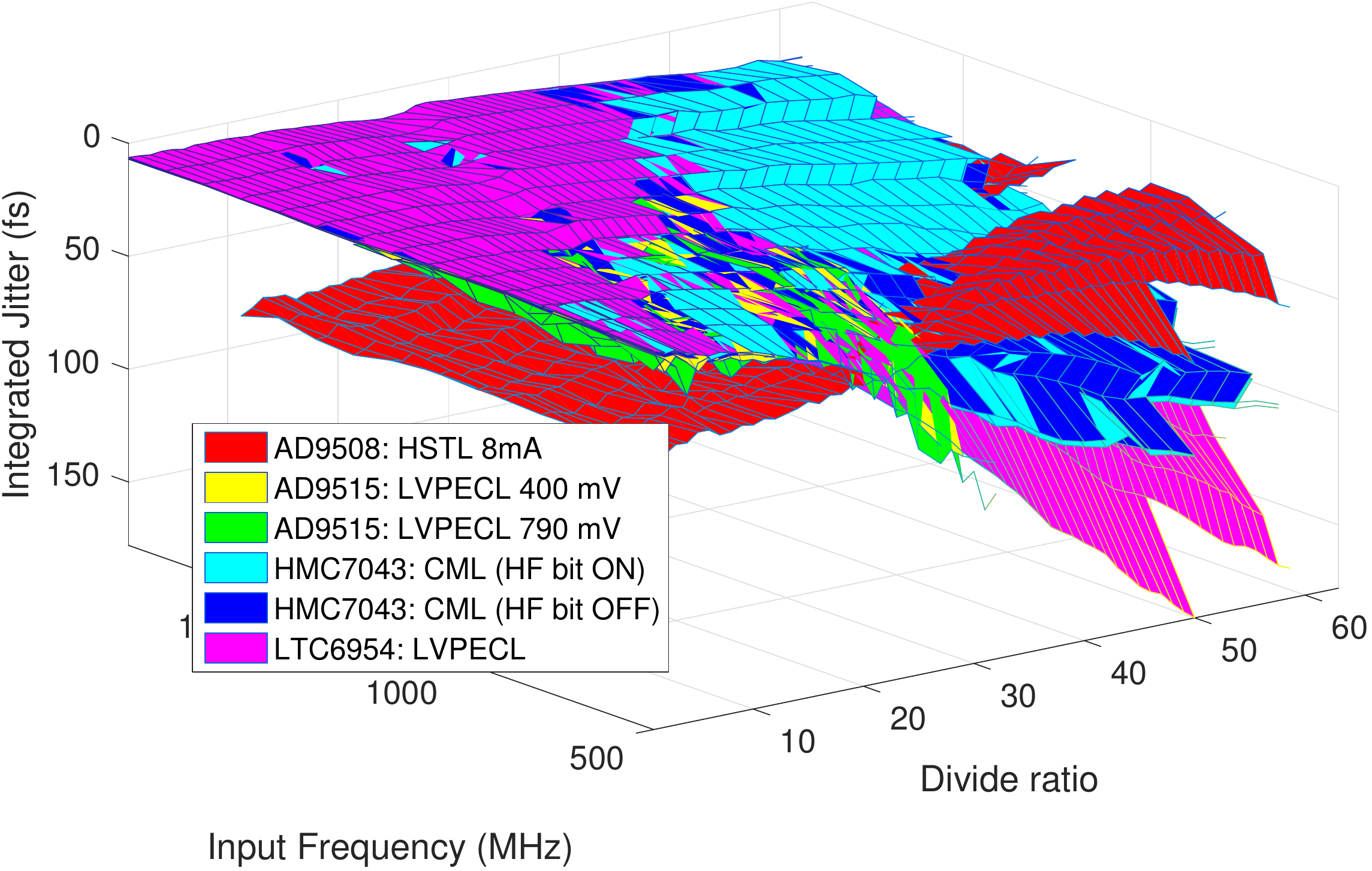}
	\caption{Close-to-carrier (100 Hz to 100 kHz) integrated phase noise (jitter) of frequency dividers as a function of the input frequency and division ratio.}
	\label{fig:f_div_comp_int_band_1e2_1e5}
\end{figure}
\begin{figure}[htb!]
	\includegraphics[width=\linewidth]{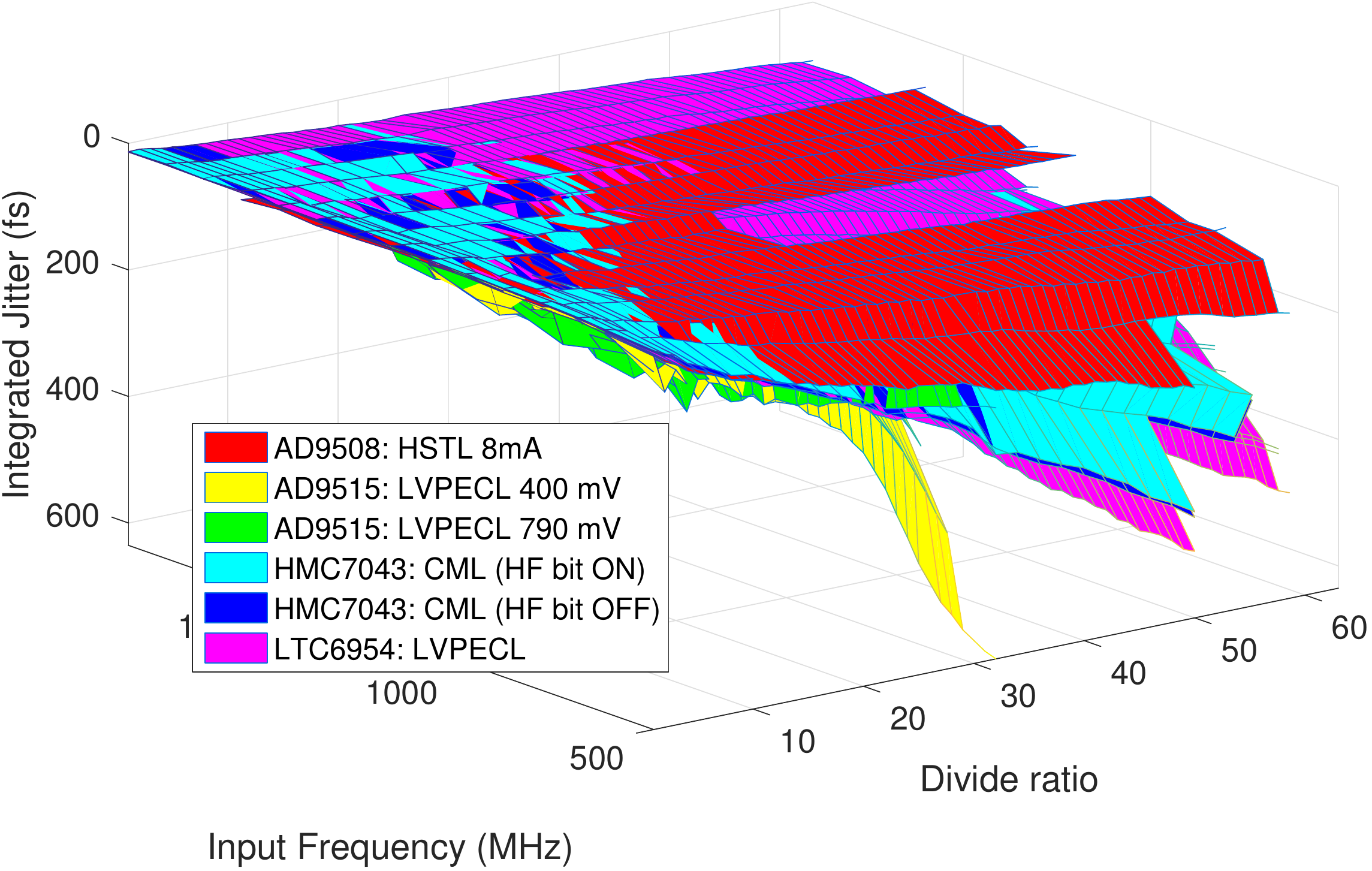}
	\caption{Far-from-carrier (100 kHz to 5 MHz) integrated phase noise (jitter) of frequency dividers as a function of the input frequency and division ratio.}
	\label{fig:f_div_comp_int_band_1e5_5e6}
\end{figure}
LTC6954 is the least sensitive of the studied frequency dividers. 

\subsubsection{Influence of reference frequency and division ratio}
Further characterization was performed with maximum safe power level at dividers' inputs.
Output phase noise was measured while the input frequency and division ratio were swept. 
The E5052B's minimum input frequency is 10 MHz, therefore the reference frequency sweep starts at 320 MHz making the characterization in 1-32 division ratios range possible.
The frequency point lists contains also 400 MHz, 500 MHz, 630 MHz (1-63 division ratios), and 12 points between 700 MHz and 1800 MHz (every 100 MHz).

To simplify the analysis the phase noise was integrated in two bands: close-to-carrier (100 Hz to 100 kHz) and far-from-carrier (100 kHz – 5 MHz). 
The bands were selected based on results of input power level sensitivity measurements.
The results are presented in Fig. \ref{fig:f_div_comp_int_band_1e2_1e5} and \ref{fig:f_div_comp_int_band_1e5_5e6}.
AD9508 offers the best jitter for high division ratios and lower input frequencies (below 1 GHz).
For high input frequencies LTC6954 and HMC7043 offers similar close-to-carrier jitter performance, which is better than other tested ICs.

\label{ss:f_div_meas}

\subsection{Passive mixers}
\label{s:passive_mixers_meas}

\begin{figure}[htb!]
	\includegraphics[width=\linewidth]{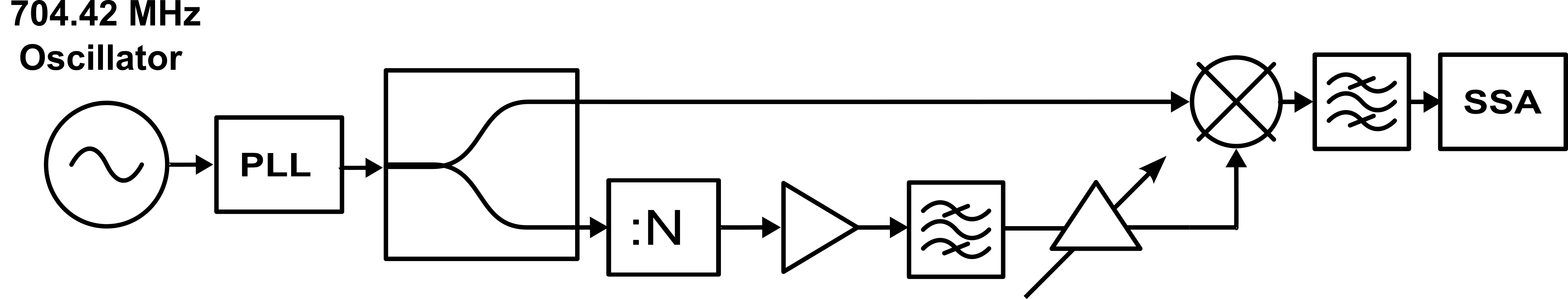}
	\caption{Test setup used to measure the influence of the IF power level on up-converting mixer output signal's phase noise.}
	\label{fig:passive_mixer_pn_test_setup}
\end{figure}

A block diagram of a test setup used to measure the influence of the IF power level on up-converting mixer output signal's phase noise in presented in Fig. \ref{fig:passive_mixer_pn_test_setup}.
Agilent E8257D generated a 704.42 MHz input signal for the custom PLL-based synthesizer, which output signal had the same frequency but improved phase noise above 50 kHz offset.
The reference signal was amplified using ADL5530 amplifier and split between a mixer (ADE-R5LH+ from MiniCircuits) and a digital frequency divider (AD9515) generating the intermediate frequency (IF) signal, which was amplified by the LHA-1H+ and low-pass filtered.
The power of the reference was kept constant at +8.1 dBm as the power of the IF signal was swept with a programmable attenuator.
Agilent E5052B measured the band-pass filtered output signal's phase noise.

\begin{figure}[htb!]
	\includegraphics[width=\linewidth]{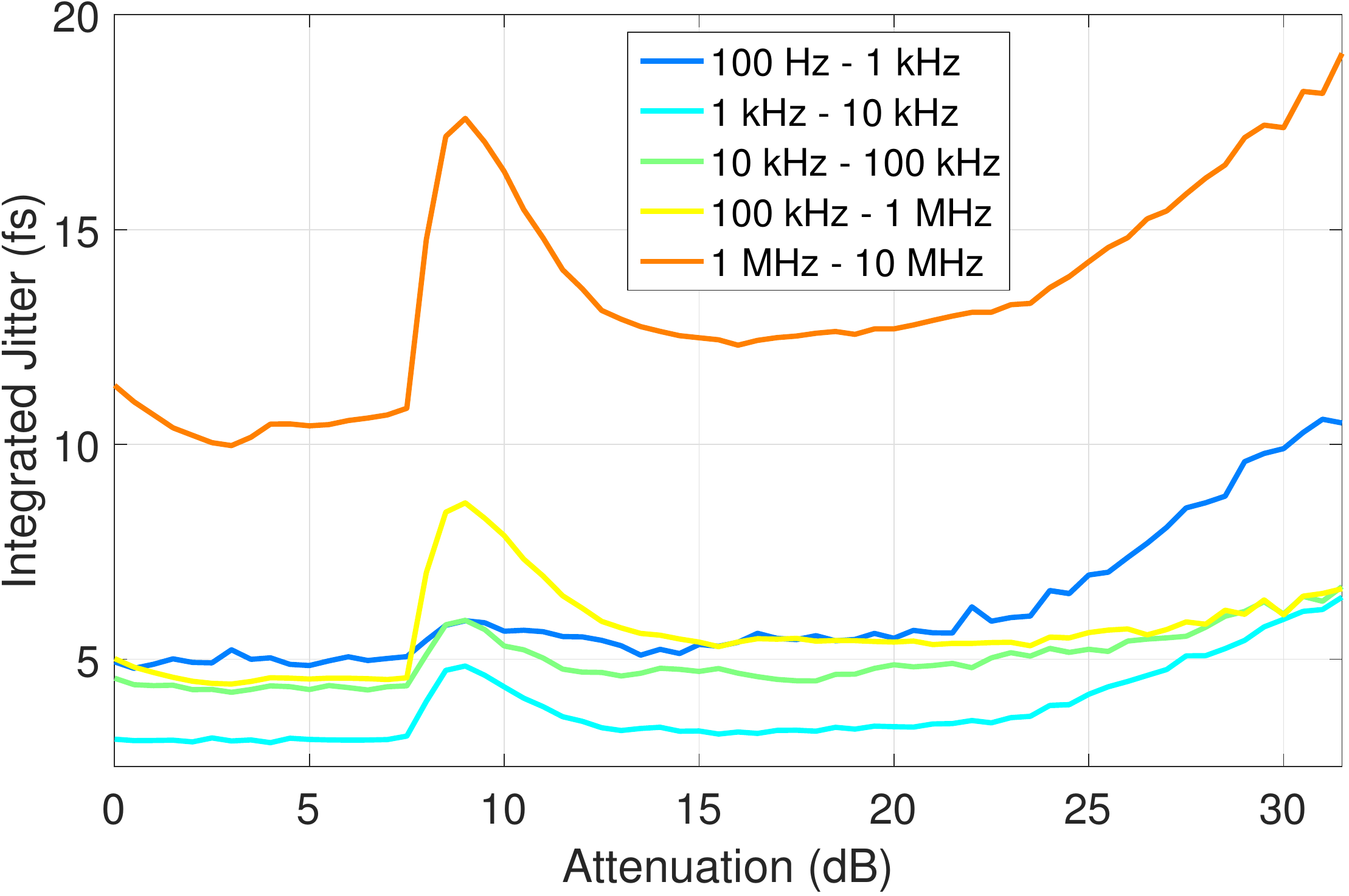}
	\caption{Integrated jitter per decade as a function of the IF signal attenuation. IF = 39.13 MHz.}
	\label{fig:lo_pn_if_power_sweep}
\end{figure}

To simplify the analysis the phase noise was integrated per offset frequency decade (Fig. \ref{fig:lo_pn_if_power_sweep}).
The IF power level has nonmonotonic influence on passive mixer output signal's phase noise.
The jitter above 10 kHz slightly drops with small attenuation (up to 3 dB).
For attenuation higher than 7 dB the jitter above 1 kHz sharply increases, peaks at 8 dB and drops utill 12 - 15 dB attenuation is reached (exact value dependent on the decade).
The far-from-carrier jitter ($\geq$ 100 kHz) rises again above the 15 dB attenuation, with all closer decades rising noticeably above 22 dB.
Similar results were obtained for slightly lower IFs (35.22 MHz, 32.02 MHz, 29.35 MHz and 27.09 MHz). 

\begin{figure}[htb!]
	\includegraphics[width=\linewidth]{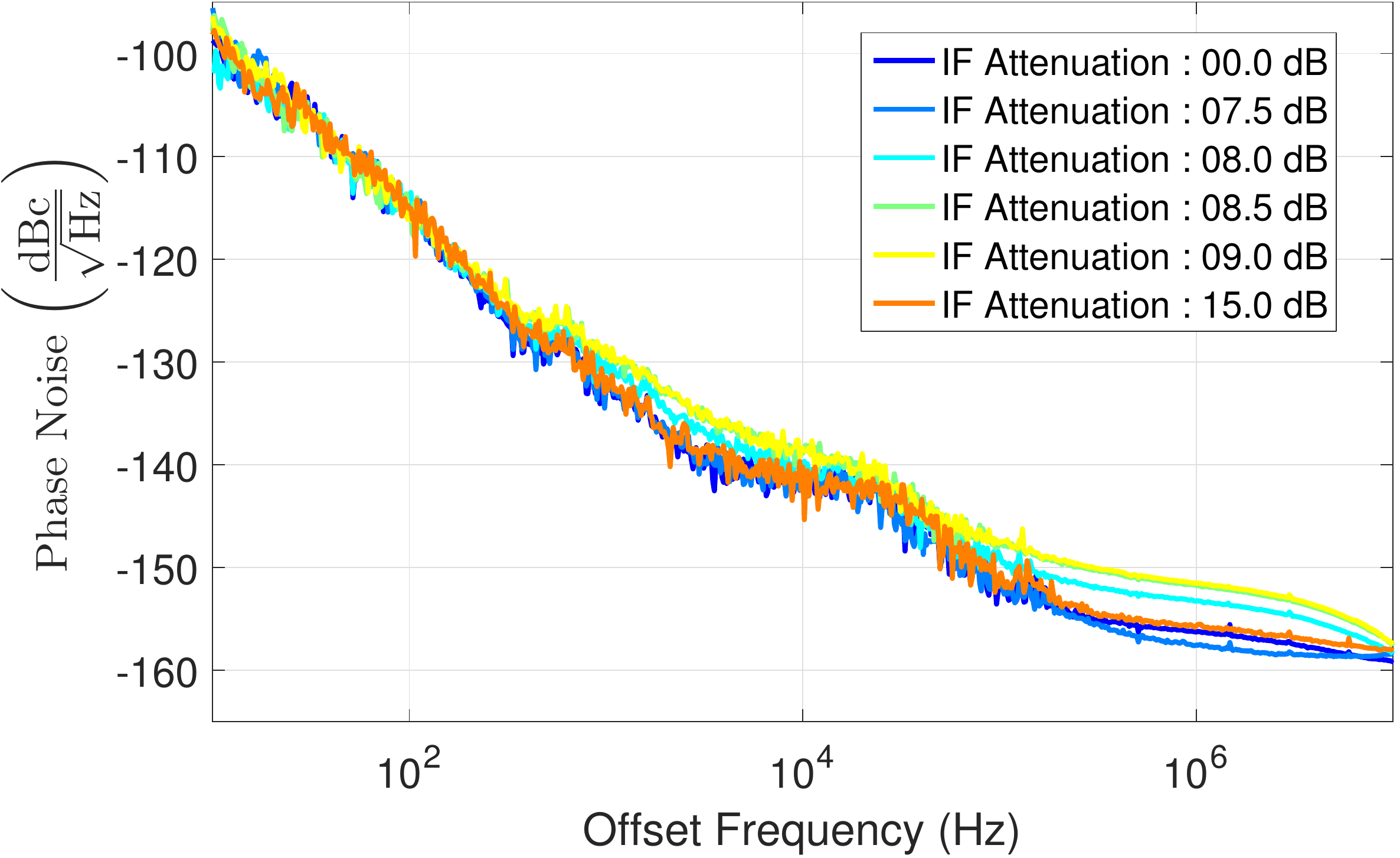}
	\caption{Phase noise of the passive mixer output signal for selected IF signal attenuation values. IF = 39.13 MHz.}
	\label{fig:lo_pn_vs_if_attenuation}
\end{figure}

\begin{figure}[htb!]
	\includegraphics[width=\linewidth]{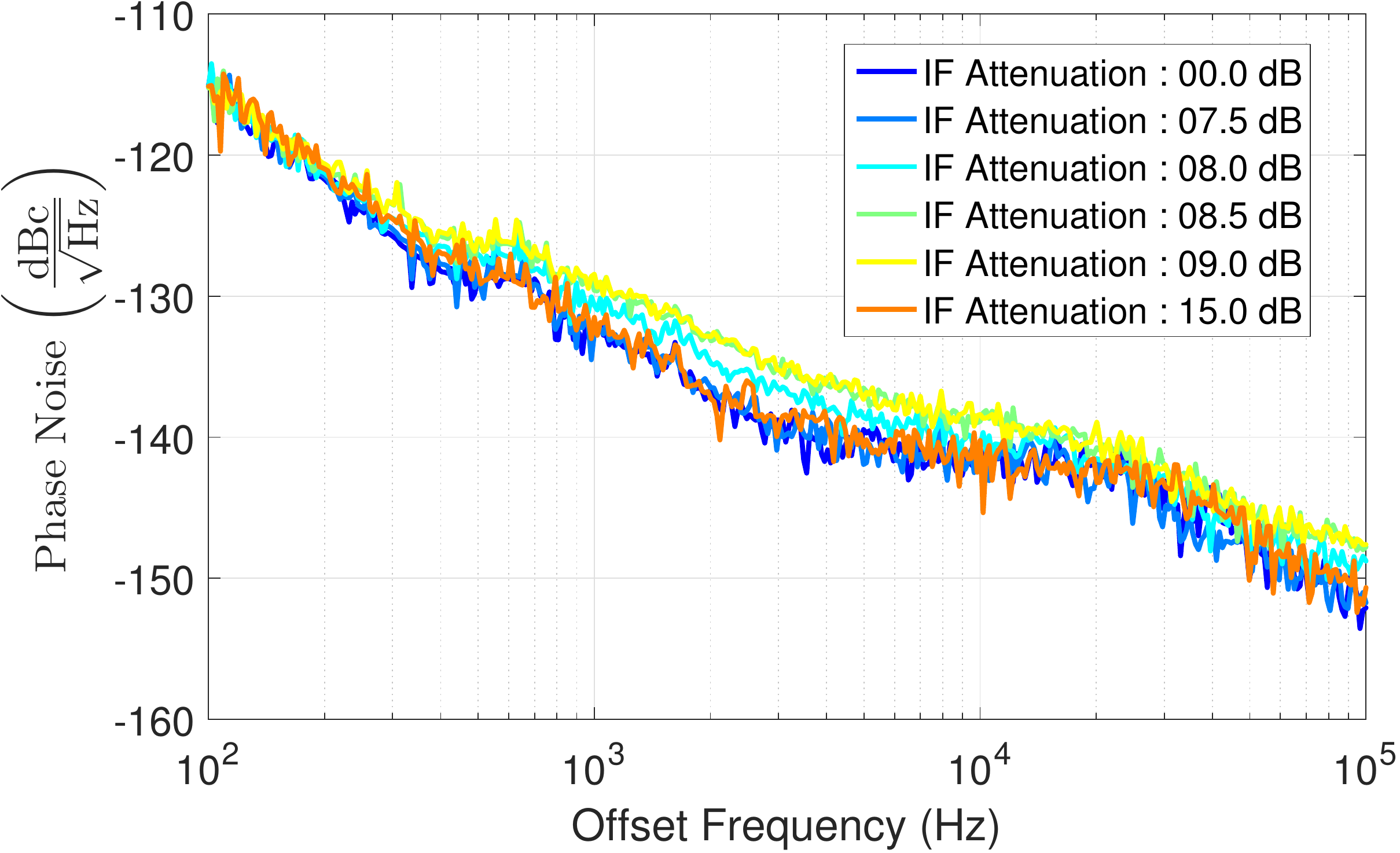}
	\caption{Phase noise of the passive mixer output signal for selected IF signal attenuation values (zoom in). IF = 39.13 MHz.}
	\label{fig:lo_pn_vs_if_attenuation_zoomin}
\end{figure}

Based on the integrated values specific attenuation values were selected for comparison of phase noise plots (Fig. \ref{fig:lo_pn_vs_if_attenuation} and \ref{fig:lo_pn_vs_if_attenuation_zoomin}).
Up to 10 Hz and 400 Hz the $1/f^2$ component coming from the reference signal dominates and in the 800 Hz - 6 kHz band flicker noise prevails.
Between 10 kHz and 100 kHz there is a bump coming from the reference signal visible, originating perhaps in the loop filter of one of the synthesisers.
Above 100 kHz the amplitude is probably constant before filtering by the band-pass filter.

\subsection{Active mixers}
\label{s:active_mixers_meas}

\begin{figure}[htb!]
	\includegraphics[width=\linewidth]{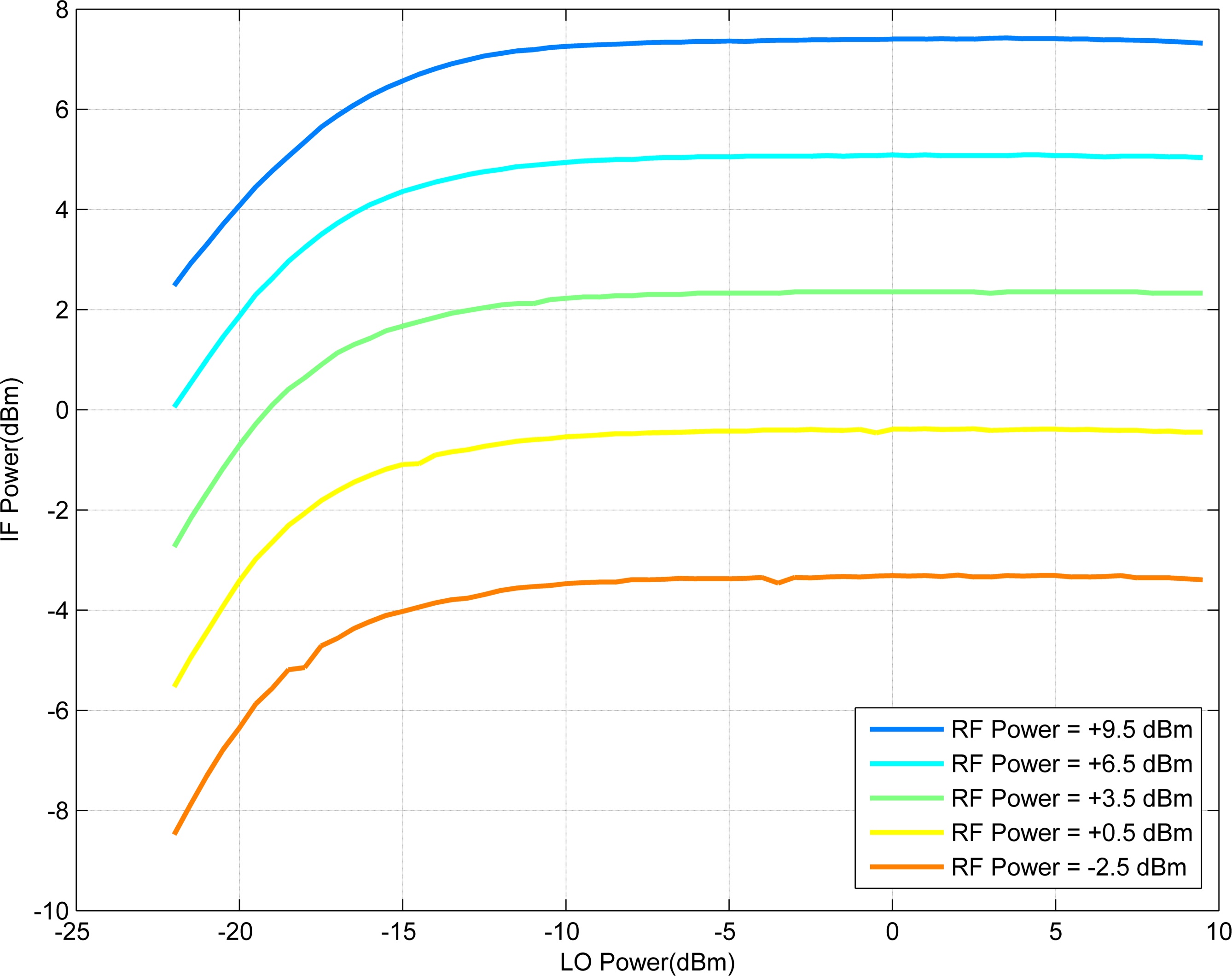}
	\caption{Transfer characteristics of an active mixer (LT5527) for selected frequencies (RF: $1300$ MHz, IF: $54\frac{1}{6}$ MHz).}
	\label{fig:LT5527_transfer_characteristic}
\end{figure}

The study of active mixer started with investigation of LT5527's transfer characteristics shown in  Fig. \ref{fig:LT5527_transfer_characteristic}.
Independent of the RF input power the output power falls as LO input power falls below -7 dBm.

Establishing the influence of the LO and RF power levels on the output signal’s integrated phase noise and amplitude noise for selected frequencies (RF: 1300 MHz, IF: $54\frac{1}{6}$ MHz) was the next step of research.
\begin{figure}[htb!]
	\includegraphics[width=\linewidth]{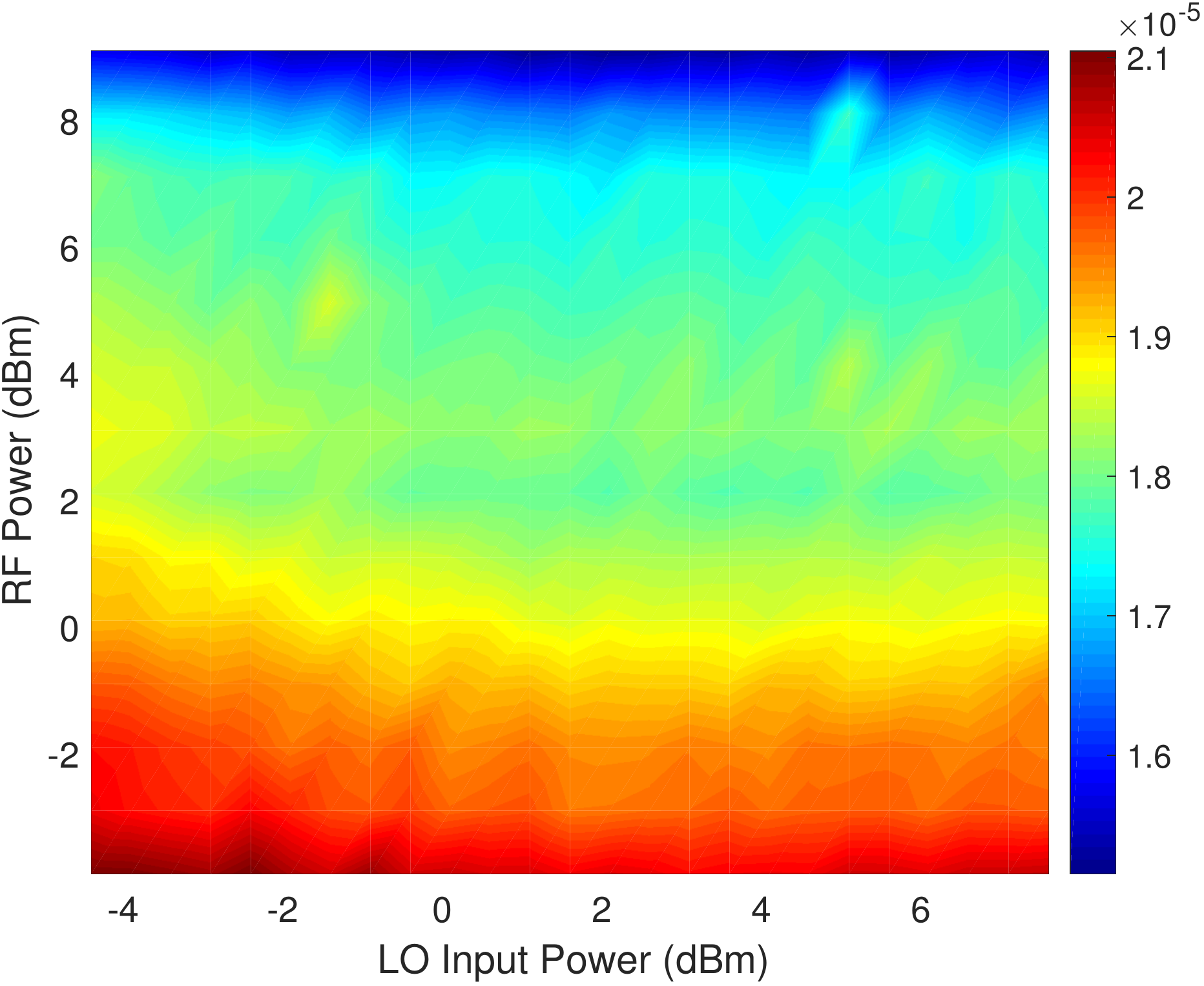}
	\caption{Amplitude noise characteristics of an active mixer's (LT5527) output signal as a functions of LO and RF power levels. Integration bandwidth: 10 Hz – 1 MHz.}
	\label{fig:LT5527_amp_jitter_pcolor}
\end{figure}
\begin{figure}[htb!]
	\includegraphics[width=\linewidth]{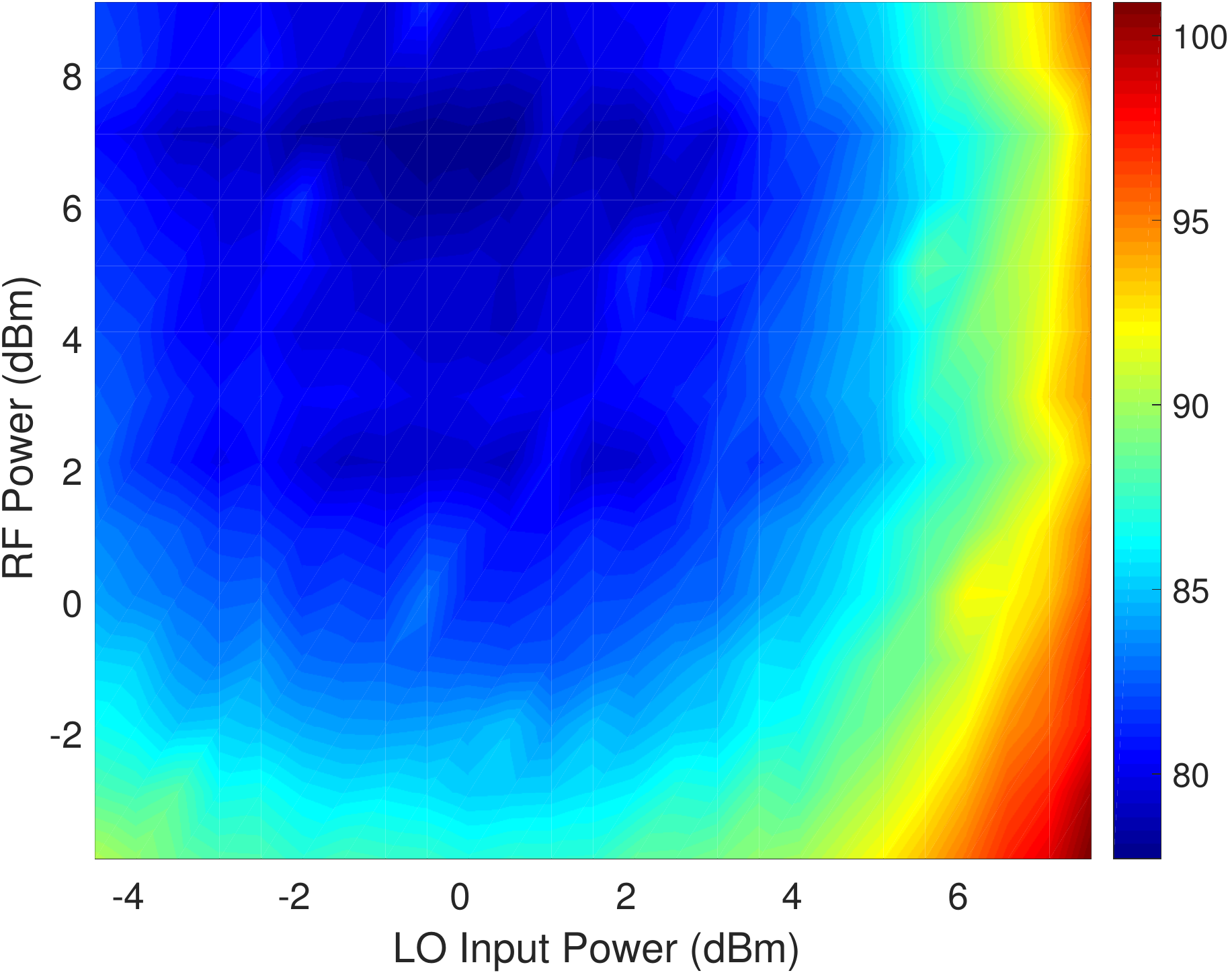}
	\caption{Phase noise characteristics of an active mixer's (LT5527) output signal as a functions of LO and RF power levels. Integration bandwidth: 10 Hz – 1 MHz.}
	\label{fig:LT5527_time_jitter_pcolor}
\end{figure}
The amplitude ripple depends mostly on the RF power and the timing (phase) jitter mostly on the LO power (Fig. \ref{fig:LT5527_amp_jitter_pcolor} and \ref{fig:LT5527_time_jitter_pcolor}). 

It was hypothesized that the increase in noise can be caused by PM-AM and AM-PM conversions of the LO signal, which can be verified by two-tone measurements.
The second tone was injected using a vector modulator module \cite{VM2}.
The modulating tone (-38 dBFS, around 120 kHz) was added to both I and Q values (the carrier, whose magnitude was constant).
the LO power was regulated using an attenuator and the RF power was constant.
When the angle between the 2nd tone and the carrier equals to 45 or 225 degrees, the output signal is purely amplitude modulated.
The output signal is purely phase modulated for phase difference of 135 or 315 degrees. 
For comparison a passive mixers (ZX05-25MH-S+) was measured in similar fashion.

\begin{figure}[htb!]
	\includegraphics[width=\linewidth]{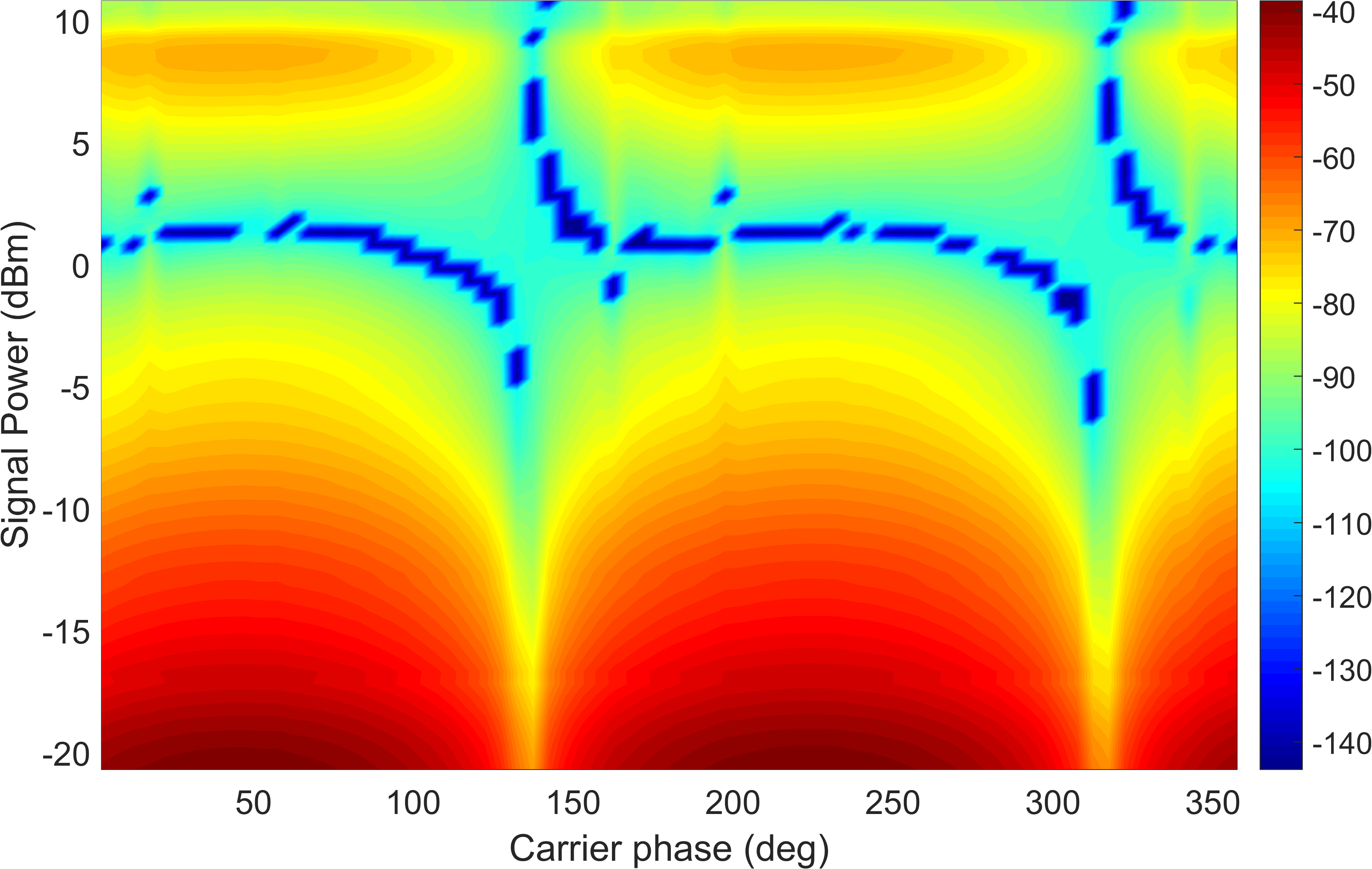}
	\caption{Amplitude modulation spur level characteristics of an active mixer's (LT5527) output signal as a functions of angle between the 2nd tone and the carrier.}
	\label{fig:LT5527_am_spur_wide}
\end{figure}

\begin{figure}[htb!]
	\includegraphics[width=\linewidth]{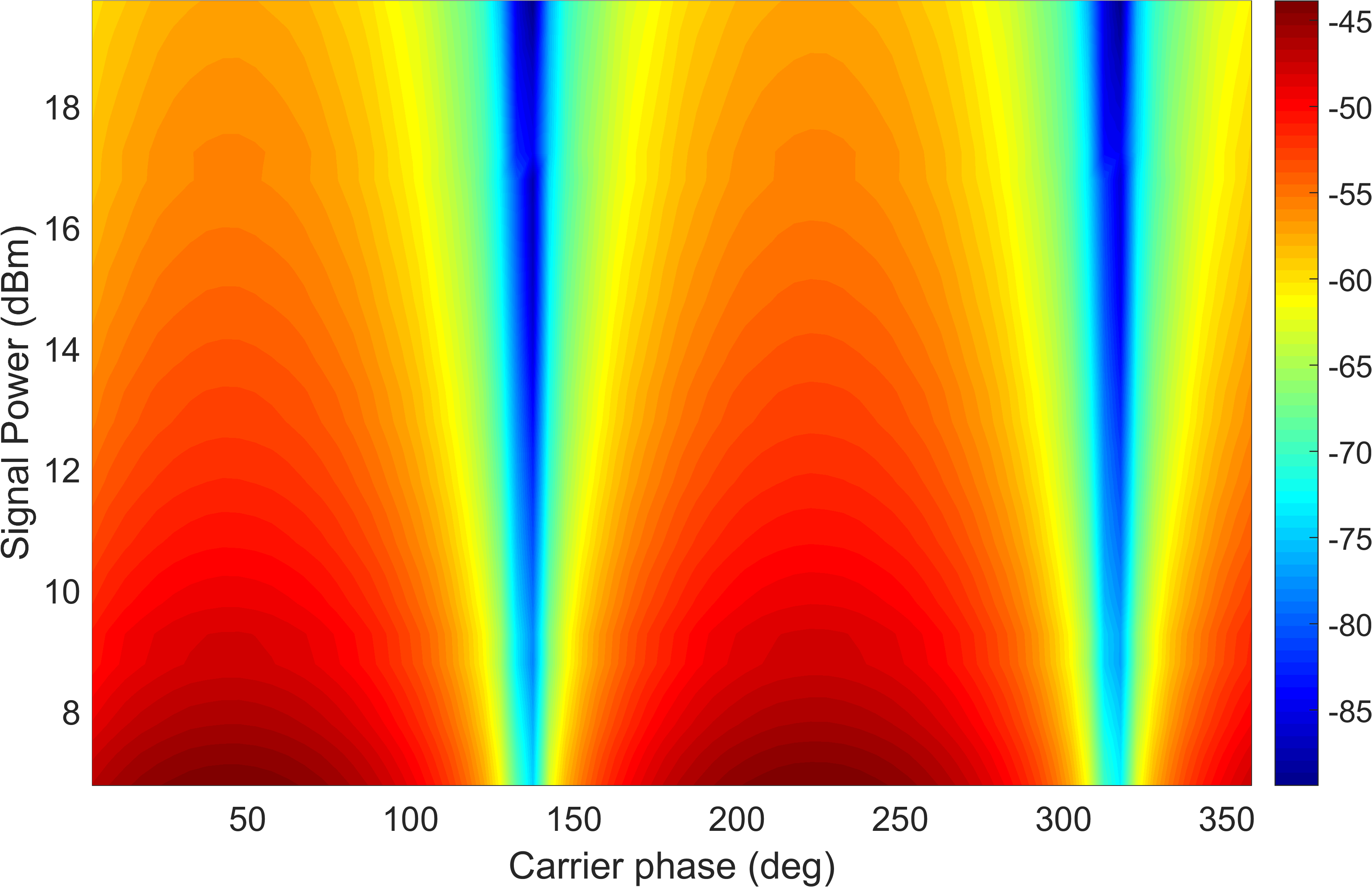}
	\caption{Amplitude modulation spur level characteristics of a passive mixer's (ZX05-25MH-S+) output signal as a functions of angle between the 2nd tone and the carrier.}
	\label{fig:ZX05_am_spur_wide}
\end{figure}

\begin{figure}[htb!]
	\includegraphics[width=\linewidth]{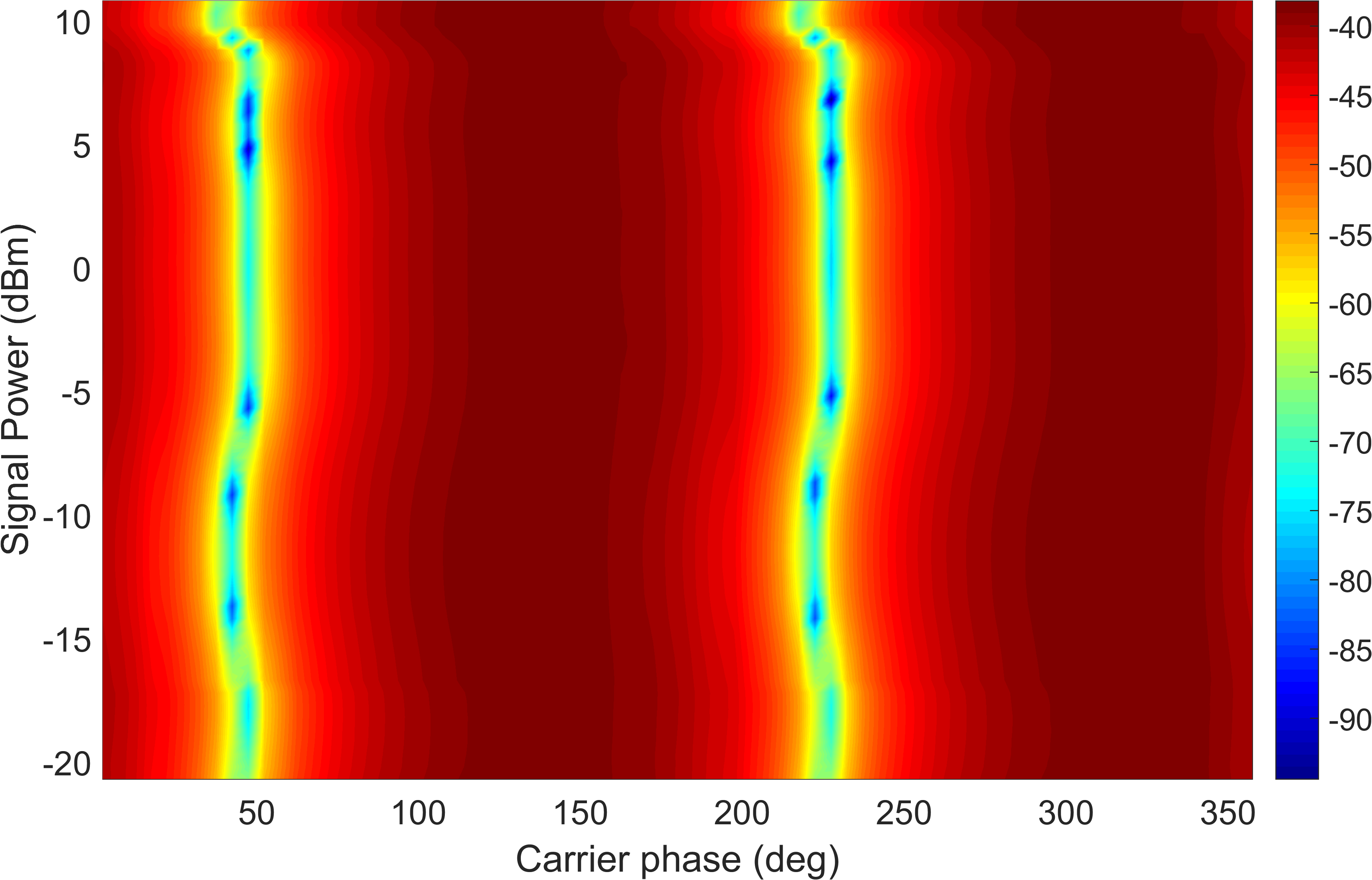}
	\caption{Phase modulation spur level characteristics of an active mixer's (LT5527) output signal as a functions of angle between the 2nd tone and the carrier.}
	\label{fig:LT5527_pn_spur_wide}
\end{figure}

\begin{figure}[htb!]
	\includegraphics[width=\linewidth]{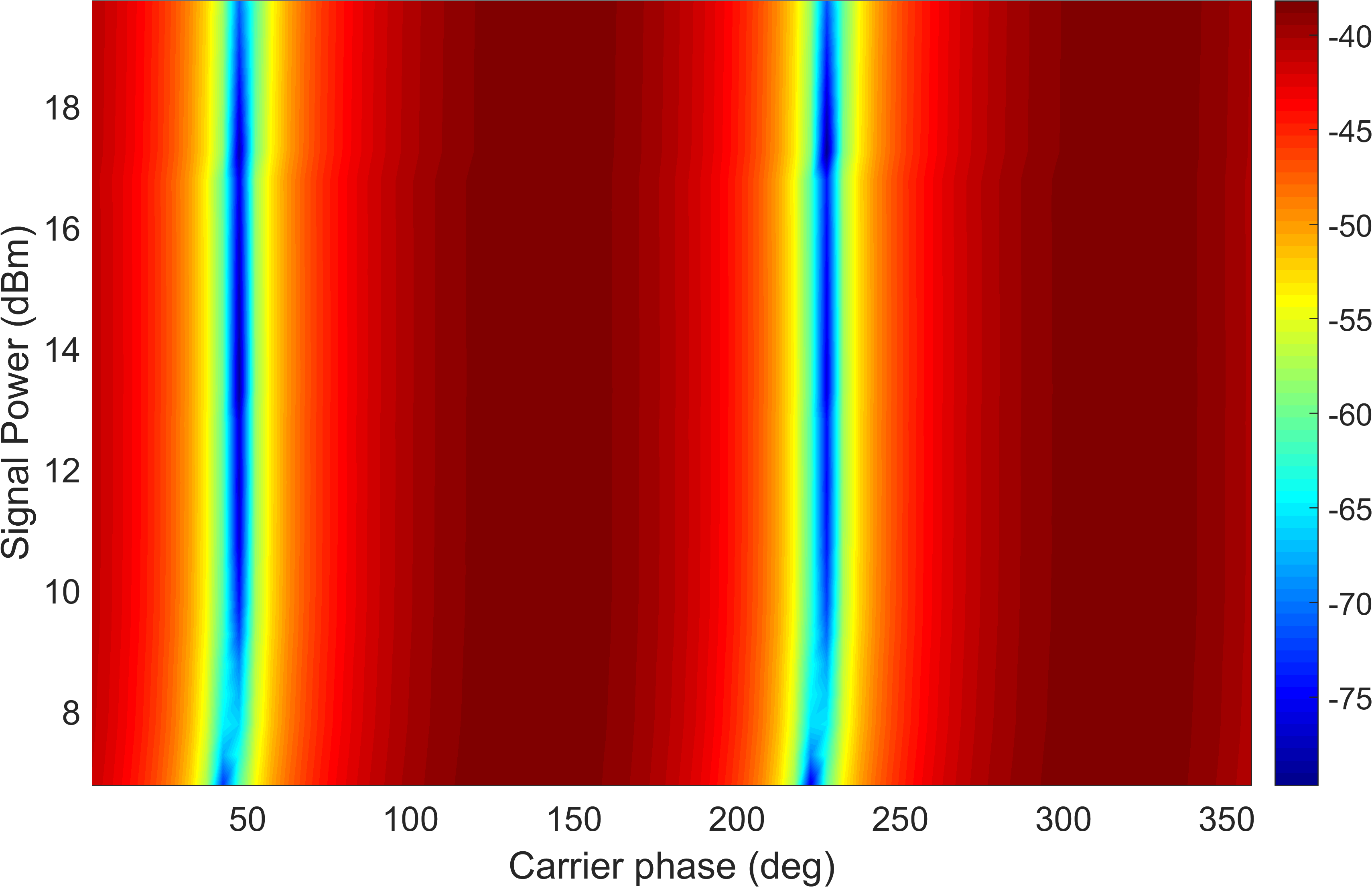}
	\caption{Phase modulation spur level characteristics of a passive mixer's (ZX05-25MH-S+) output signal as a functions of angle between the 2nd tone and the carrier.}
	\label{fig:ZX05_pn_spur_wide}
\end{figure}

As Fig. \ref{fig:LT5527_am_spur_wide} shows, in case of the active mixer the amplitude spur level is minimal for the input power around +2 dBm.
The passive mixer shows very different behaviour, the suppression of the amplitude spur at its output increases monotonically with the input power (Fig. \ref{fig:ZX05_am_spur_wide}).
In both mixers, independent of the input power, the phase spur is strong ($\ge$-50 dBc) outside the narrow range centred at pure phase modulation (Fig. \ref{fig:LT5527_pn_spur_wide} and \ref{fig:ZX05_pn_spur_wide}).
These results suggest that modulation conversions are probably not responsible for increase in active mixer output signal's noise.
The findings might not be applicable to other ICs.

\section{Conclusion}
\label{s:conclusion}
This paper began with a summary of frequency synthesis schemas commonly used in LLRF control systems and of the current state of research on additive phase noise in selected frequency converting devices.
Selected COTS frequency dividers ICs were characterized.
Nonmonotonic influence of the IF power level on passive mixer output signal's phase noise was observed.
The reason for increase of active mixers output signal's phase noise with the LO input power was not clearly established.

\section*{ACKNOWLEDGMENT}
Work supported by Polish Ministry of Science and Higher Education, decision number DIR/WK/2016/03.

\bibliographystyle{IEEEtran}

\begin{thebibliography}{1}

\bibitem{Kroupa-1982}
V. F.~Kroupa ``Noise Properties of PLL Systems,"  in {\it IEEE Trans. Comm.}, vol. COM-30, no. 10, pp. 2244--2252, Oct 1982.

\bibitem{Phillips-1987}
D. E.~Phillips ``Random Noise in Digital Gates and Dividers,"  in {\it IEEE 41st Annu. Symp. Freq. Control}, pp. 507--511, July 1987.

\bibitem{Egan-1990}
W. F.~Egan ``Modeling Phase Noise in Frequency Dividers,''  in {\it IEEE Trans. Ultrason., Ferroelect., Freq. Control}, vol. 37, no. 4, pp. 307--315, July 1990.

\bibitem{Levantino-2004}
S. Levantino, L. Romano, S. Pellerano, C. Samori and A. L. Lacaita, ``Phase Noise in Digital Frequency Dividers,''  in {\it IEEE Journal Solid-state Circuits}, vol. 39, no. 5, pp. 775--784, May 2004.

\bibitem{Apostolidou-2008}
M. Apostolidou, Peter G.M.~Baltus and Cicero S. Vaucher, ``Phase Noise in Frequency Divider Circuits'' in {\it 2008 IEEE Int. Symp. Circuits Syst.}, pp. 2538--2541, May 2008.

\bibitem{Zuk-2016}
M. Żukociński, K. Czuba, Ł. Zembala, U. Mavrič, M. Hoffmann and F. Ludwig, ``Low Phase Noise Local Oscillator and Clock Generation for Cavity Field Detection,'' in {\it Proc. 20th IEEE-NPSS Real Time Conf.}, Padua, Italy, June 2016.


\bibitem{Barnes}
C. A. Barnes, A. Hati, C. W. Nelson and D. A. Howe, "Residual PM noise evaluation of radio frequency mixers," {\it. 2011 Joint Conf. IEEE Int. Freq. Control and European Freq. Time Forum Proc.}, pp. 1--5, San Fransisco, CA, 2011.

\bibitem{Rubiola}
Enrico Rubiola ``Tutorial on the double balanced mixer,'' August 2006.

\bibitem{Faber}
Marek T.~Faber, Jerzy Chramiec and Mirosław E.~Adamski ``Microwave and millimeter-wave diode frequency multipliers,'' Artech House, 1995.

\bibitem{Gan-2016}
N. Gan, R. Liu, X. P. Ma and Y. L. Chi,``Design and Evaluation of a Broad Band microTCA.4 Based Downconverter,'' in {\it Proc. 7th Int. Particle Accelerator Conf.}, pp. 2746--2748, June 2016.

\bibitem{Hoffmann-PhD}
Matthias Hoffmann ``Development of a multichannel RF field detector for the Low-Level RF control of the Free-Electron Laser at Hamburg'', Ph.D. dissertation, Hamburg University of Technology, 2008.

\bibitem{Hull-1993}
C. D. Hull and R. G. Meyer, "A systematic approach to the analysis of noise in mixers," in {\it IEEE Trans. Circuits and Syst. I: Fundamental Theory and Applications}, vol. 40, no. 12, pp. 909-919, Dec 1993.

\bibitem{Rudell-1997}
J. C. Rudell, J. J. Ou, T. B. Cho, G. Chien, F. Brianti, J. A. Weldon and P. R. Gray, ``A 1.9-GHz wide-band IF double conversion CMOS receiver for cordless telephone applications'', in {\it IEEE Journal of Solid-State Circuits}, vol. 32, no. 12, pp.~2071--2088, December 1997.

\bibitem{Rofougaran-1996}
A. Rofougaran, J. Y. C. Chang, M. Rofougaran and A. A. Abidi, ``A~1~GHz~CMOS RF front-end IC for a direct-conversion wireless receiver'', in {\it IEEE Journal of Solid-State Circuits}, vol. 31, no. 7, pp.~880--889, July 1996.

\bibitem{Terrovitis-1999}
M. T. Terrovitis and R. G. Meyer, ``Noise in current-commutating CMOS mixers,'' in {\it IEEE Journal of Solid-State Circuits}, vol. 34, no.~6, pp. 772--783, June 1999.

\bibitem{Darabi-2000}
H. Darabi and A. A. Abidi, ``Noise in RF-CMOS mixers: a simple physical model'' in {\it IEEE Journal of Solid-State Circuits}, vol. 35, no.~1, pp. 15--25, January 2000.

\bibitem{LT5527_DS}
Linear Technology Corporation ``LT5527 Datasheet,'' Revision A. 

\bibitem{VM2}
I.~Rutkowski, et al., K. Czuba, M. Grzegrzólka, D. Makowski, A. Mielczarek, P. Perek and H. Schlarb, ``Improved Vector Modulator Card for MTCA-based LLRF Control System for Linear Accelerators," in {\it Proc. Int. Particle Accelerator Conf.}, Shanghai, China 2013.

\end{thebibliography}

\end{document}